\documentclass{amsart}

\usepackage{t1enc,amsthm,amssymb,array,epsfig}

\newtheorem{lema}{Lemma}
\newtheorem{prop}[lema]{Proposition}
\newtheorem{teo}[lema]{Theorem}
\newtheorem{rem}[lema]{Remark}

\theoremstyle{definition}
\newtheorem{defin}{Definition}

\newcommand{\R}{\ensuremath{\mathbb{R}}}
\DeclareMathOperator{\Bola}{B}
\newcommand{\bola}[2]{\Bola_{#2}(#1)}
\DeclareMathOperator{\Vol}{Vol}
\newcommand{\vol}[1]{\Vol(#1)}
\newcommand{\schr}{Schr\"odinger}
\newcommand{\dis}{\displaystyle}
\newcommand{\Gd}{\mathcal{G}_d}
\newcommand{\cd}{\ensuremath{c_d}}
\newcommand{\Nclass}{\ensuremath{\mathcal{N}_\gamma(\Gd;\R^n)}}
\DeclareMathOperator{\capac}{cap}
\DeclareMathOperator{\trace}{tr}
\DeclareMathOperator{\tr}{tr}

\begin{document}
\date{\today}

\title[Intrinsic Moment of Inertia of Membranes and Yang-Mills Theories]{Intrinsic Moment of Inertia of Membranes as bounds for the mass gap of Yang-Mills Theories.}
\author[M.P. Garcia del Moral, L. Navarro, A. J. P\'erez A. and A. Restuccia]{M.P. Garcia del Moral$^1$, L. Navarro $^2$, A. J. P\'erez A.$^3$ and A. Restuccia$^4$}
\address{$^{1}$  Perimeter Institute for Theoretical Physics, Waterloo, Canada, Ontario N@J 2Y5, Canada \\
Dept. of Physics and Astronomy, MacMaster University,
1280 Main Street West,Hamilton,Ontario,L8S 4M1,Canada \\
DAMTP, DAMTP, Centre for Mathematical Sciences, University of
Cambridge CB3 0WA, UK.}
\email{ mmoral@perimeterinstitute.ca, M.G.d.Moral@damtp.cam.ac.uk}
\address{$^{2}\,$ Departamento de Matem\'aticas, Universidad Sim\'on Bol\'\i
var, Apartado 89000, Caracas 1080-A, Venezuela.}
\email{lnavarro@ma.usb.ve}
\address{$^{3}\,$ Departamento de Matem\'aticas, Universidad Sim\'on Bol\'\i
var, Apartado 89000, Caracas 1080-A, Venezuela.}
\email{ajperez@ma.usb.ve}
\address{$^{4}$ Departamento de F\'\i sica, Universidad Sim\'on Bol\'\i
var, Apartado 89000, Caracas 1080-A, Venezuela. }
\email{arestu@usb.ve} \maketitle

\begin{abstract} 
We obtain the precise condition on the potentials of Yang-Mills
theories in $0+1$ dimensions and $D0$ brane quantum mechanics
ensuring the discretness of the spectrum. It is given in terms of
a moment of inertia of the membrane. From it we obtain a bound for
the mass gap of any $D+1$ Yang-Mills theory in the slow-mode
regime. In particular we analyze the physical case $D=3$. The
quantum mechanical behavior of the theories, concerning its
spectrum, is determined by harmonic oscillators with frequencies
given by the inertial tensor of the membrane. We find a class of
quantum mechanic potential polynomials of any degree, with
classical instabilities that at quantum level have purely discrete
spectrum.
\end{abstract}


\vskip 1 cm
\section{Introduction}
The characterization of the discretness of bosonic potentials in
quantum mechanics represents a very challenging problem by itself
and in relation to matrix model regularizations of field
theories and to dimensional reduction of Yang-Mills to
$0+1$ dimensions. It would provide information about the
spectrum of quantum non-abelian gauge theories, as well as
theories containing gravity (bosonic membrane of M-theory).

The potential of the 2-brane theories is quartic on the fields and
exactly the same, after a $SU(N)$ regularization, to the
dimensional reduction of Yang-Mills theories to 1 dimension
(time). The potential in terms of the membrane fields has valleys
extending to infinity along the directions of zero potential.
There are configurations, which have been called string-like spikes
\cite{dwln, helling} in the context of $D=11$ supermembrane
theory, for which the energy is zero. They can be attached to the
membrane and even connect two membranes without any change in
the energy of the system. Those configurations correspond from the
Yang-Mills point of view to gauge fields valued on abelian
subalgebras of $SU(N)$, in particular on the Cartan subalgebra. A
toy model for these class of potentials was first studied in
\cite{simons} and then extended to a toy model of the $D=11$
supersymmetric membrane in \cite{dwln}. The potentials are
classically unstable, quantum mechanically stable and its
supersymmetric extension quantum unstable. Although the bosonic
potential presents flat directions extending to infinity the walls
of the valleys become narrower as we move to infinity in a way
that the wave function cannot escape to infinity. A proof of
discretness of the spectrum of the $SU(N)$ regularized hamiltonian of the bosonic
membrane was presented in \cite{lucher}, where a bound
\begin{equation}
\langle \Psi,H\Psi\rangle\geq\langle \Psi,\lambda\Psi\rangle
\end{equation}
in terms of a  function $\lambda(x)$, with $\lambda(x)\to\infty$
as $\|x\|\to\infty$ was obtained. It is a sufficient condition for
the discretness of the spectrum. However in order to understand
the exact property of the membrane, and, as a consequence of
Yang-Mills, ensuring the discretness of the spectrum one should
look for a necessary and sufficient condition. That condition was
discovered by A.M. Molchanov, \cite{molchanov} and recently
extended by Maz'ya and Shubin \cite{Maz-Shu}, and makes use of the
mean value, in the sense of Molchanov, of the potential on an
$n$-dimensional cube. If the mean value goes to infinity at large
distances in configuration space the spectrum is discrete,
otherwise is continuous. We consider the Molchanov condition for
the membrane potential. By performing the explicit calculation we
obtain that at large distances in the direction of the valleys,
the mean value of the potential behaves as an harmonic oscillator
with frequencies determined by a moment of inertia of the
membrane. There is a inertial tensor associated to the potential
of the membrane, or equivalently of the $D0$ mechanics, which is
strictly positive. One may interpret the mean value of the
potential as a rotational energy of $D0$ branes.

This is the precise property ensuring the discretness of the
membrane and of Yang-Mills in $0+1$ dimensions. This approximation for Yang-Mills describe the
spectrum of the theory when only the modes with zero momenta but non-zero energy
are kept. This regime has been thought to contain relevant
information of the IR phase of the theory in which the gluons are
expected to be confined forming bound states called glueballs.

We then obtain bounds for the Yang-Mills hamiltonian in terms of
the moment of inertia of the membrane. The hamiltonian bound implies that the resolvent of the Schr\"odinger operator is compact. It also provide a bound for the mass gap. The same results are proven for the Yang-Mills theories in the slow mode regime and give an strong indication that the complete Yang-Mills
theories in D-dimensions have a mass gap.\newline

The interesting properties of the membrane and Yang-Mills
potentials are also present in more general potentials,  of higher
degree than four. We discuss a class of these potentials in
section 3, in particular they are related to the $D=11$ $5$-brane
potential. We prove discretness of these class of potentials.
\newline

 The
paper is organized as follows: In section 2 some preliminary
results are presented. In section 3 we obtain a class of
potentials for which the spectrum of the Schr\"odinger operator is discrete,
extending previous results. In section 4 we introduce the moment
of inertia of the membrane In section 5 we obtain a lower bound
to the hamiltonian in terms of it. In section 6. the discretness
of the Yang-Mills hamiltonian in the slow mode regime is proved.
In section 7 we present the conclusions.

\section{Preliminary Results}
In 1953 A. M. Molchanov (see \cite{molchanov}) proved that in dimension $n=1$, the discretness of the spectrum of the  Schr\"odinger operator $-\Delta+V$ in $L^2(\R^n)$ with a locally integrable and semi-bounded below potential $V$  is equivalent to
\[\textrm{For every fixed}\quad d>0\quad\int_{Q_d}V(x)\,dx\to\infty\quad \textrm{as}\quad Q_d\to\infty\]
where $Q_d$ is an open cube with the edges parallel to coordinate axes and of edge length $d$  and $Q_d\to\infty$ means that the distance from $Q_d$ to 0 goes to infinity. In the same paper Molchanov showed that this condition is not sufficient for dimension $n\ge2$, his counter-example is based on theorem 2.1 of his paper:
\begin{quotation}\emph{
The spectrum of the  Schr\"odinger operator $-\Delta+V$  is not discret if and only if there exist a constant $C>0$ and a sequence of disjoint cubes $\{Q_k\}_{k=1}^\infty$ with the same edge length such that for all $k=1,2,\ldots$
\[\lambda(Q_k)=\inf\int_{Q_k}(|\nabla\psi(x)|^2+V(x)|\psi(x)|^2)\,dx<C\]
where the infimum is taken over all the functions $\psi$ such that $\psi|_{\partial Q_k}=0$ and $\dis\int_{Q_k}|\psi(x)|^2\,dx=1$.}
\end{quotation}
After, Molchanov stated and proved a modified necesary an sufficient condition for the discretness of the spectrum of Schr\"odinger operators for $n\ge2$, which involve the class of negligible compact subsets of the cube. We will only state the V. Maz'ya and M. Shubin's generalization (see \cite{Maz-Shu}). Before we state this theorem, let us remember the following definition of capacity. For the details and the full-general version of the theorem see \cite{Maz-Shu}.
\begin{defin}
Let $n\ge3$, $F\subset\R^n$ be compact, and $Lip_{c}(\R^n)$ the set of all real-value functions with compact support satisfying a uniform Lipschitz condition in $\R^n$. Then the Wiener's capacity of $F$ is defined by
\[\capac(F)=\capac_{\R^n}(F)=\inf\left\{\int_{\R^n}|\nabla u(x)|^2\,dx\Big|\ u\in Lip_{c}(\R^n),\ u|_F=1\right\}\] 
\end{defin}
In physical terms the capacity of the set $F\subset\R^n$ is defined as the
electrostatic energy over $\mathbb{R}^{n}$ when the electrostatic
potential is set to $1$ on $F$.

\begin{defin}
Let $\Gd\subset\R^n$ be an open, bounded and star-shaped set of diameter $d$, let $\gamma\in(0,1)$. The \emph{negligibility class} \Nclass\
consists of all compact sets $F\subset\overline{\Gd}$ satisfying
$\capac(F)\leq\gamma\capac(\overline{\Gd})$.
\end{defin}
Balls and cubes in $\R^n$ are useful examples of such $\Gd$. In what follows we denote the ball of
diameter $d$ and center $x$ by $\Bola_{d}(x)$ and the $n-$dimensional Lebesgue measure by $\vol{\cdot}$.

\begin{teo}[Maz'ya and Shubin]\label{teo_mazya}
Let $V\in L^1_{\text{loc}}(\R^n)$, $V\geq0$.

Necessity: If the spectrum of $-\Delta+V$ in $L^2(\R^n)$ is discrete
then for every function $\gamma:(0,+\infty)\rightarrow(0,1)$ and
every $d>0$
\begin{equation}\label{inf_int}
\inf_{F\in\Nclass}\int_{\Gd\setminus
F}V(x)\,dx\rightarrow+\infty\quad\text{as}\quad
\Gd\rightarrow\infty.
\end{equation}

Sufficiency: Let a function $d\mapsto\gamma(d)\in(0,1)$ be defined for $d>0$ in a neighborhood of $0$ and satisfying
\[ \limsup_{d\downarrow0}d^{-2}\gamma(d)=+\infty. \]

Assume that there exists $d_0>0$ such that
\textrm{(\ref{inf_int})} holds for every $d\in(0,d_0)$. Then the
spectrum of $-\Delta+V$ in $L^2(\R^n)$ is discrete.
\end{teo}

\begin{rem}
It follows from the previous theorem that a necessary condition
for the discreteness of spectrum of $-\Delta+V$ is
\begin{equation}\label{necessary}
\int_{\Gd}V(x)\,dx\rightarrow\infty
\quad\text{as}\quad\Gd\rightarrow\infty.
\end{equation}
\end{rem}

Let's recall that in 1934, K. Friedrichs (see \cite{Maz-Shu} for further references) proved that the spectrum of the
\schr\ operator $-\Delta+V$ in $L^2(\R^n)$ with a locally
integrable potential $V$ is discrete provided
$V(x)\rightarrow\infty$ as $|x|\to\infty$.

\begin{rem}
The presence of negligibility sets $F$ in the formulation of the Maz'ya and Shubin's theorem is related to the first term in the expression of $\lambda(Q_k)$. In fact this term describes also an electrostatic energy, when $\Psi=1$ on $F\subset Q$ the energy is minimized by the harmonic solution to the corresponding Dirichlet problem. The energy of the harmonic solution defines the capacity of the capacitor $[F,Q]$ while the capacity of $F$ is the capacity of the capacitor $[F,\infty]$.
\end{rem}

The following lemma is very usefull tool in  the next sections.
\begin{lema}\label{lema}
For each given $\Gd=\Gd(x_0)$,
\[ \cd:=\inf_{F\in\Nclass}\vol{\Gd\setminus F}>0. \]
\end{lema}
\begin{proof}
Let $V(x)=|x|$. Then by Friedrichs theorem the spectrum of
$-\Delta+V$ is discrete, so by theorem \ref{teo_mazya} we have
\[ \inf_{F\in\Nclass}\int_{\Gd\setminus F}V(x)\,dx\to\infty\quad\text{as}\quad |x_0|\to\infty. \]
Now $\int_{\Gd\setminus F}V(x)\,dx\leq(|x_0|+d)\vol{\Gd\setminus
F}$ implies that
\[ \inf_{F\in\Nclass}\int_{\Gd\setminus F}V(x)\,dx \leq (|x_0|+d)\inf_{F\in\Nclass}\vol{\Gd\setminus F}, \]
from which follows that $\inf_{F\in\Nclass}\vol{\Gd\setminus
F}>0$, as we claimed.
\end{proof}

\section{Some potentials for which the spectrum of its \schr\ operator is discrete}
In this section we use the sufficiency conditions for the
discreteness of spectrum provided by theorem \ref{teo_mazya}.

\begin{teo}
Let $V\geq0$, $V\in L^1_{\text{loc}}(\R^n)$ and $\cd$ as in
lemma~\ref{lema}. Suppose that there is some $d_0>0$ such that
for all $d\in(0,d_0)$, there exists an open neighborhood $\Omega_d$
of $V^{-1}(0)$ with the following properties:
\begin{enumerate}
\item[(i)] In $\R^n\setminus\Omega_d$, $V(x)\to\infty$ as $|x|\to\infty$.
\item[(ii)] $\vol{\Bola_d\cap\Omega_d}<\frac{\cd}{2}$ for every ball $\bola{x_0}{d}$ with $|x_0|>>d_0$.
\end{enumerate}
Then the spectrum of the \schr\ operator $-\Delta+V$ in
$L^2(\R^n)$ is discrete.
\end{teo}
\begin{proof}
Let $\Gd=\bola{x_0}{d}$ with $d\in(0,d_0)$ fixed. Let
$F\in\Nclass$ and $E=(\Gd\setminus F)\setminus(\Gd\cap\Omega_d)$.
Notice that $E\subset\R^n\setminus\Omega_d$. By condition (i) for
each $M>0$, there exists $R>0$ such that
\[\textrm{If }x\in\R^n\setminus\Gd\textrm{ and }|x|>R, \textrm{ then }V(x)>M. \]

In this way, if $|x_0|-d>R$ then $|x|> R$ for all $x\in\Gd$, and as
$V\geq0$ we have
\[ \int_{\Gd\setminus F}V(x)\,dx \geq \int_E V(x)\,dx \geq M\vol{E}\geq M\left(\vol{\Gd\setminus F}-\frac{\cd}{2}\right) \]
and by the lemma~\ref{lema}
\[ \inf_{F\in\Nclass}\int_{\Gd\setminus F}V(x)\,dx\geq M\frac{\cd}{2}\quad\text{if $|x_0|>R+d$}. \]
This together with theorem \ref{teo_mazya} implies that the
spectrum of $-\Delta+V$ is discrete.
\end{proof}

Now, let us apply the previous theorem.
\begin{prop}
Let $V(x)=\dis\prod_{k=1}^n|x_k|^{\alpha_k}$, where $\alpha_k>0$
for all $k=1,2,\ldots,n$. Then the spectrum of the \schr\ operator
$-\Delta+V$ in $L^2(\R^n)$ is discrete.
\end{prop}
\begin{proof}
Let $d_0=\frac{n2^n}{\omega_n}$, where $\omega_n$ denotes the area
of the unit ball, and let $d\in(0,d_0)$ fixed. Consider
$U_k=\{x\in\R^n\;|\; |x_k|<\epsilon_d\}$, where $0<\epsilon_d<1$
will be determined below. Let $\Omega_d=\dis\cup_{k=1}^nU_k$. Then
$V^{-1}(0)\subset\Omega_d$.
\begin{enumerate}
\item[(a)] In $\R^n\setminus\Omega_d$, $V(x)\to\infty$ as $|x|\to\infty$:

First, $x\in\R^n\setminus\Omega_d$ implies $|x_k|\geq\epsilon_d$
for all $k=1,2,\ldots, n$.

Let $|x_j|=\dis\max_{k=1,\ldots,n}|x_k|$,
$\alpha_M=\dis\max_{k=1,\ldots,n}\alpha_k$ and
$\alpha_m=\dis\min_{k=1,\ldots,n}\alpha_k$. Then, denoting
$|\alpha|=\alpha_1+\cdots+\alpha_n$ we have
\[ V(x) = \prod_{k=1}^n|x_k|^{\alpha_k} \geq \epsilon_d^{|\alpha|-\alpha_j}\left(\max_{k=1,\ldots,n}|x_k|\right)^{\alpha_j} \geq \epsilon_d^{|\alpha|-\alpha_m}\left(\frac{|x|}{\sqrt{n}}\right)^{\alpha_j}. \]

Thus, for $|x|>1$ we have
\[ V(x)\geq \frac{\epsilon_d^{|\alpha|-\alpha_m}}{(\sqrt{n})^{\alpha_M}}|x|^{\alpha_m}, \]
from which follows that $V(x)\to\infty$ as $|x|\to\infty$.

\item[(b)] $\vol{\Bola_d\cap\Omega_d}<\frac{\cd}{2}$ for every ball $\bola{x_0}{d}$ with $|x_0|>>d_0$:

For $|x_0|$ large enough, if
$\bola{x_0}{d}\cap\Omega_d\neq\emptyset$ then $\bola{x_0}{d}\cap
U_k\neq\emptyset$ for only one $k\in\{1,2,\ldots,n\}$, namely
$k=j$. In consequence,
\[ \bola{x_0}{d}\cap\Omega_d\subset\{x\in\R^n\;|\;|x_j|<\epsilon_d,\: |x_k-x_{0j}|<d,\ k=1,\ldots,j-1,j+1,\ldots,n\},\]
from which follows that $\vol{\bola{x_0}{d}\cap\Omega_d}\leq
2^n\epsilon_d d^{n-1}$ and this, in turn, implies what we wanted
to prove if we choose
\begin{equation}\label{epsilon_choice}
\epsilon_d<\frac{\cd}{2^nd^{n-1}}.
\end{equation}
\end{enumerate}

Thus, by the previous theorem we conclude that the spectrum of
$-\Delta+V$ is discrete.
\end{proof}
\begin{rem}
Given that $\cd\leq\frac{\omega_nd^n}{n}$ then
inequality~(\ref{epsilon_choice}) implies that
$\epsilon_d<\frac{\omega_nd}{2^{n}n}$, which tends to zero as
$d\to0$ and also
$\epsilon_d<\frac{\omega_nd}{2^{n}n}<\frac{\omega_nd_0}{2^nn} =
\frac{\omega_n}{2^nn}\frac{n2^n}{\omega_n}=1$.
\end{rem}

\section{Moment of Inertia and Discretness of the potential for membrane theories}

The matrix regularization of the bosonic membrane in the light
cone gauge was done long time ago by
\cite{dwnh,hoppe,helling,halpern}. It consists in expanding the
bosonic fields $X^{m}$ in an orthonormal basis
$Z_{A}(\sigma)$, defined on the Riemann surface $\Sigma$,
\begin{equation}
X_{m}=X_{m}^{A}(\tau)Z_{A}(\sigma).
\end{equation}
The structure constants arising from the symplectic bracket on
$\Sigma$, $g_{AB}^{C}$, are those of the infinite group of area
preserving diffeomorphisms.

The idea of matrix regularization is to integrate the dependence
on the spatial coordinates and to regularize this model by an
$SU(N)$ one, in such a way that we end up with a theory of quantum
mechanics of matrices, whose structure constants $f_{AB}^{C}$
converge in the large $N$ limit to the original ones $g_{AB}^{C}$,
recovering the original symmetries in the large $N$ limit. This
limit is in general not rigorously known, and a discussion on it
will be presented in a separate paper \cite{l1}. In this paper we
are only concerned with the theory at finite $N$. The potential
associated to the membrane is described by
\begin{equation}
V=(Y^{A}_{m}Y^{B}_{n}f_{AB}^C)^{2},
\end{equation}
where $Y_m,\ m=1,..,d$, denote the transverse direction to the light
cone coordinates $Y^{+}Y^{-}$. $A,B,C$ denote $SU(N)$ indices and
the
 structure constants,
\begin{equation}
f_{AB}^C=-2iN\sin\left(\frac{A\wedge B }{N}\pi\right)\delta_{A+B}^C,
\end{equation}
$A=(a_{1},a_{2})\quad a_{1},a_{2}=1,..,N-1$ modulo
 $N$ and with
$|m|+|n|\neq 0$.
 The structure constants in the large $N$ limit are proportional
 to the structure constants of the area preserving
 diffeomorphisms, the residual gauge symmetry of the membranes in
 the light cone gauge. The large $N$ limit is defined by taking
 indices $A,B$ on a fixed set $\Lambda$ and taking $N\to\infty$.
We are interested in the evaluation of
\[ \inf_{F\in\Nclass}\int_{\Gd\setminus F}V,\]
where $\Gd$ is the star-shaped set and $F$ the negligible set
satisfying $\capac F\leq\gamma \capac\Gd$ defined in section 1, as the
distance of $\Gd$, to a fixed point in $\mathbb{R}^{n}$goes to
infinity. $V$ satisfy the assumptions of theorem 1. The
integration is performed with the Lebesgue measure $dY$. We
introduce for a given $F\in\Nclass$, the center of mass of the
volume $\Gd\setminus F$ by
\begin{equation}
\int_{\Gd\ F}Y^{A}_{m}dY =X^{A}_{m}\vol{G/ F}. \end{equation} We
decompose,
\begin{equation}
Y^{A}_{m}=X^{A}_{m}+x^{A}_{m},
\end{equation}
with
\begin{equation}
\int_{\Gd\setminus
F}x^{A}_{m}dY=\int_{[\Gd\setminus
F]_{0}}x^{A}_{m}dx=0
 \end{equation}
 where $[\Gd\setminus
F]_{0}$ denote the set with its center of mass at the origin of
coordinates. We then have
\begin{equation}
V(Y)=\sum_{m}\overline{(X_{m}^{A}+x^{A}_{m})}D_{AB}^{m}(X_{m}^{B}+x_{m}^{B}),
 \end{equation}
 where
\begin{align}
&D_{AB}^{m}\equiv\sum_{n\neq
m}(X_{n}^{C}+x_{n}^{C})\overline{(X_{n}^{D}+x_{n}^{D})}f_{BC}^{E}\overline{f}_{AD}^{E}=\\
\nonumber & =4N^{2}\sum_{n\neq
m}(X_{n}^{C}+x_{n}^{C})\overline{(X_{n}^{D}+x_{n}^{D})}\sin\left(\frac{B\wedge
C }{N}\pi\right)\sin\left(\frac{A\wedge D}{N}\pi\right)\delta_{B+D}^{A+C},
 \end{align}
 and
 \begin{equation}
V(Y)=\sum_{m}[(\overline{X}_{m}^{A}D_{AB}^{m}X_{m}^{B}+
O(X_{m})]. \end{equation} Here $\lim_{\|X_{m}\|\to
\infty}\frac{O(X_{m})}{X_{m}}=0$ assuming the
other coordinates of the center of mass bounded. We will consider
first the case when only one $X_{m}$ tends to infinity, namely $X_{m_0}\to\infty$, while all the others $X_{m}, m\neq
m_0$ remain bounded.
 We get
\begin{align}\label{I}
&\int_{\Gd\setminus F}V(Y)dY=\int_{[\Gd\setminus F]_{0}}V(X+x)dx=
\sum_{m}\left[(\overline{X}_{m}^{A})\left(\int_{[\Gd\ F]_{0}}D_{AB}^{m}dx\right)X^{B}_{m}+O(X_{m})\right]=\\
\nonumber
&=\sum_{m}\overline{X}_{m}^{A}\left[\mathbb{I}_{AB}^{m}+4N^{2}\sum_{n\neq
m} X_{n}^{C}\overline{X}_{n}^{D}\sin\left(\frac{A\wedge
C}{N}\pi\right)\sin\left(\frac{B\wedge D}{N}\pi\right)\delta_{B+D}^{A+C}\vol{\Gd\
F}\right]X_{m}^{B}\\ \nonumber
&+O(X_{m}),
\end{align}
where
\begin{align}
\mathbb{I}_{AB}^{m}=\sum_{n\neq m}4N^{2}\int_{[\Gd\setminus
F]_{0}}x_{n}^{C}\overline{x}_{n}^{D}\sin\left(\frac{A\wedge
C}{N}\pi\right)\sin\left(\frac{B\wedge D\pi}{N}\right)\delta_{A+D}^{B+C}\, dx
\end{align}
is an intrinsic inertial tensor of the membrane associated to the
distribution $[\Gd\setminus F]_{0}$. $\mathbb{I}_{AB}^{m}$ is a
self-adjoint positive matrix. The diagonal terms are
\begin{align}
\mathbb{I}_{AA}^{m}=4N^{2}\int_{[\Gd\setminus F]_{0}}\sum_{n\neq
m}x_{n}^{C}\overline{x}_{n}^{C}\sin^{2}\left(\frac{A\wedge
C}{N}\pi\right)\, dx.
\end{align}
Its trace is
\begin{align}
\trace{\mathbb{I}^{m}}=\sum_{A}\mathbb{I}_{AA}^{m}=2N^{4}\int_{[\Gd\setminus
F]_{0}}\sum_{n\neq m}x_{n}^{C}\,\overline{x}_{n}^{C}\,dx.
\end{align}
This is the moment of inertia of the mass distribution
$\Gd\setminus F$ with respect to the subspace generated by
$X_{m}^{A}$.
 The second term in equation (\ref{I}) is quartic on the center
of mass coordinates. Its contribution to the integral of the
potential is the value of the potential at the center of mass
times the volume of the $\Gd\setminus F$,
\begin{align}
(X_{m}^{A}X_{n}^{C}f_{AC}^{E})^{2}\vol{\Gd\setminus F}.
\end{align}
This term becomes zero when we take the center of mass at the
bottom of the valleys, in the directions where $V$ is zero. We
then conclude that along these directions where the potential is
zero we obtain,
\begin{align}\label{II}
&\lim_{\|X_{m_0}\|\to \infty}\left(\inf_{F\in\Nclass}\int_{\Gd\setminus
F}V\,dx\right)= \lim_{\|X_{m_0}\|\to
\infty}\left(\inf_{F\in\Nclass}\overline{X}_{m_0}^{A}\mathbb{I}_{AB}^{m_0}X_{m_0}^{B}\right)
\\ \nonumber &
=\left(\lim_{\|X_{m_0}\|\to\infty}\|X_{m_0}\|^2\right)\inf_{F\in\Nclass}\left(\frac{\overline{X}_{m_0}^{A}}{\|X_{m_0}\|}\mathbb{I}_{AB}^{m_0}\frac{X_{m_0}^{B}}{\|X_{m_0}\|}\right),
\end{align}
and we notice that
\[\inf_{F\in\Nclass}\left(\frac{\overline{X}_{m_0}^{A}}{\|X_{m_0}\|}\mathbb{I}_{AB}^{m_0}\frac{X_{m_0}^{B}}{\|X_{m_0}\|}\right)\ge\inf_{F\in\Nclass}\lambda_{F},
\]
where $\lambda_{F}$ is the minimum eigenvalue of $\mathbb{I}_{AB}^{m_0}$. It may be shown by explicit calculations and using  lemma~\ref{lema} that the infimum of the right hand side of last inequality is greather than 0. One may also obtain that conclusion in an indirect way from theorem 1. In fact 
we know from an independent argument (see next section or
\cite{lucher}), that the spectrum of the Schr\"odinger operator
we are considering is discrete with finite multiplicity. According
to theorem 1, the left hand member of equation
(\ref{II}) must then go to infinity when $\|X_{m_0}\|\to\infty$,
hence it is neccesary that
\[\inf_{F\in\Nclass}\lambda_{F}>0.\]

If the center of mass goes to infinity in more than one direction $X_m,\ m=1,\ldots, d$, (19) remains valid. The inertial tensor has now an additional index $\mathbb{I}_{AB}^{mn}$ where $m,n=1,\ldots, d$. And it is a  strictly positive self adjoint matrix, the diagonal terms have the same expression (16). The corresponding equality (19) holds when the center of mass of $\Gd\setminus F$
is located on the directions where the potential is zero. In other
directions where the potential does not vanish the left hand side
of (\ref{II}) is greater or equal to the quadratic right hand side
of (\ref{II}).

We remark that this relation is the most precise one can obtain for
the quantum Schr\"odinger operator, since the condition of the
theorem in section 1 is a necessary and sufficient condition. The
mean value, in the sense of Molchanov, of the membrane potential
is bounded by the mean value of an harmonic oscillator at large
distances in the configuration space and it is equal to it at the
directions of zero potential. This bound implies, using Maz'ya and Shubin's theorem, that the corresponding Schr\"odinger operator is discrete with finite multiplicity.

\section{Bounds for the Hamiltonian}
We present in this section a bound for the hamiltonian of the
regularized bosonic M2 brane in terms of the integrand of the
moment of inertia we discussed in the previous section. It is an
independent argument showing the discretness of the spectrum and
providing bounds to all the eigenvalues of the membrane. It
represents a generalization of the proof given in \cite{simons}.

Hereafter we use the notation $x_m^A$ instead of $Y_m^A$ for the configuration variables.
We re-arrange the potential of the M2 brane in the form
\begin{align}
V(x)=\sum_{m}^{d}x^{A}_{m}\overline{x}^{B}_{m}M_{AB}^{m}
\end{align}
where
\begin{align}
M_{AB}^{m}=\sum_{n\neq
m}f_{AE}^{C}x^{E}_{n}\overline{f}_{BD}^{C}\overline{x}_{n}^{D}
\end{align}
is a positive hermitian matrix. Then, the hamiltonian can be
written as,
\begin{align}
&H=-\frac{1}{2}\Delta+\frac{1}{2}(-\Delta+2V),\\
\nonumber &-\Delta+2V=\sum_{m}\left[-\left(\frac{\partial}{\partial
x^{A}_{m}}\right)^{2}+2x^{A}_{m}\overline{x}^{B}_{m}M_{AB}^{m}\right]\geq\sum\sqrt{2}(\trace{M^{m}})^{1/2},
\end{align}
hence
\begin{align}
 H\geq\frac{1}{2}\left(-\Delta +\sqrt{2}\sum_{m}(\tr M^{m})^{1/2}\right).
\end{align}
The computation for $\trace{M^{m}}$ gives
\begin{align}
\tr M^{m}=2N^{4}\sum_{n\neq m}x^{E}_{n}\overline{x}^{E}_{n}.
\end{align}
We finally get the operatorial bound,
\begin{align}
H\geq\frac{1}{2}\left(-\Delta+2N^{2}\sum_{m}\left(\sum_{m\neq
n}(x^{E}_{n}\overline{x}^{E}_{n})^{1/2}\right)\right)\equiv\widetilde{H}.
\end{align}
The potential $\widetilde{V}\equiv 2N^{2}\sum_{m}(\sum_{m\neq
n}x^{E}_{n}\overline{x}^{E}_{n})^{1/2}$, satisfies the criteria:
$\widetilde{V}\to\infty$ when
$\|x\|=(\sum_{n}x_{n}^{E}x_{n}^{E})^{1/2}\to\infty$ implying that
the resolvent of $H$ is compact, $H$ has a discrete spectrum with
finite multiplicity. The formula also provides a bound for the
mass gap of $H$ given by the first eigenvalue of the operator
$\widetilde{H}$.\\

\section{Spectrum of Yang-Mills Theories in the slow-mode regime}
In this section we will consider the explicit relation between
dimensional reduced Yang-Mills theories to $0+1$ dimensions and
the membrane moment of inertia condition ensuring the discretness
of the spectrum .

The hamiltonian of Yang-Mills in $D+1$ space-time dimensions has
the following hamiltonian
\begin{align}
H_{ym}=\int_{\Sigma_{2}}
\frac{g^{2}}{2}(\Pi^{i})^{2}+\frac{1}{4g^{2}}F_{ij}^{a}F_{ij}^{a}-
\mathcal{D}_{i}\Pi^{ai}A_{0}^{ai}
\end{align}
where $\Pi^{ai}$ represents the conjugate momenta to $A_{i}^{a}$.
The indices of color are $a=1,\dots, \dim SU(N)=N^{2}-1$, and
$i=1,\dots,D$ labels the spatial components of gauge fields. The
last piece of the Hamiltonian represents the Gauss law:
\begin{align}
\mathcal{D}_{i}\Pi^{ai}=0.
\end{align}
We  assume we are in the slow mode regime and all of the fast
modes has been integrated out in such a way that we deal with an
effective hamiltonian containing gauge fields $A_{i}(t)$ without
spatial dependence. This approximation has been considered to
keep relevant information of the theory in the IR limit where the
gluons are supposed to be confined forming glueballs bound states.
The hamiltonian has then the expression,
\begin{align}
H_{ym}^{red}=\tr\left[\frac{g^{2}V_{m}}{2}(\Pi^{i})^{2}+\frac{V_{m}}{4g^{2}}(f^{abc}A_{i}^{b}A_{j}^{c})^{2}-
[A_{i},\Pi^{i}]^{a}A_{0}^{ia}\right](t).
\end{align}
In order to make comparison with the hamiltonian of the preceding
sections we redefine the fields in terms of dimensionless ones.
Following \cite{gabadadze} we take,
\begin{align}
A_{i}\to \frac{g^{1/3}}{V_{m}^{1/3}}A_{i},\quad P^{i}\to
\frac{1}{g^{2/3}V_{i}^{2/3}}P_{i}.
\end{align}
The hamiltonian
can be re-written as,
\begin{align}
H_{ym}^{red}=\frac{g^{2/3}}{V_{m}^{1/3}}\tr\left[\frac{1}{2}(\Pi^{i})^{2}+\frac{1}{4}(f^{abc}A_{i}^{b}A_{j}^{c})^{2}-
[A_{i},\Pi^{i}]^{a}A_{0}^{ia}\right](t)=\frac{g^{2/3}}{V_{m}^{1/3}}H.
\end{align}
The mass operator of the regularized  bosonic membrane also in
terms of dimensionless variables can be written as,
\begin{align}
\frac{M^{2}}{2}=T^{2/3}\tr\left[\frac{1}{2}(\Pi^{i})^{2}+\frac{1}{4}(f^{abc}A_{i}^{b}A_{j}^{c})^{2}-
[A_{i},\Pi^{i}]^{a}A_{0}^{ia}\right](t)=T^{2/3}H.
\end{align}
 This means that the two theories are equivalent and the
discretness of Yang-Mills in the slow mode regime is related to
the existence of a tensor of moment of inertia of the membrane. We
interpret it as an strong evidence that the configuration space
for Yang-Mills theories in the large $N$ limit is compact. The
discretness condition then is realized as a condition of
"rotational energy" of $D0$ branes in the configuration space of
the bosonic membrane.
\subsection{Mass Gap and analytic eigenvalues distribution}
An interesting problem is to estimate the mass gap of the theory
for a particular value of $N$. We have from the previous section
\begin{align}
H\geq\frac{1}{2}\left(-\Delta+2N^{2}\sum_{m}\left(\sum_{m\neq
n}(x^{E}_{n}\overline{x}^{E}_{n})^{1/2}\right)\right)\equiv\widetilde{H}.
\end{align}

Let us now consider, as an example, the case of $SU(3)$ in $D=3+1$
since it is the gauge group realized in the nature. The number of
degrees of freedom once the gauge fixing condition has been
taken into account is $d=2$, $A=1,\dots,8$, and $N=3$.
$\widetilde{H}$ is bounded from below by the following hamiltonian

\begin{align}
H_{1}=-\Delta+2N^{2}(X^{I}X^{I})^{1/2} \quad I,J=1,\dots 16
\end{align}
The eigenfunctions of $H_{1}$ are expressed in terms of Bessel
functions. The eigenvalues of $H_{1}$ are lower bounds, one by
one, of the eigenvalues of $H$. That is, $SU(3)$ Yang-Mills in the
slow-mode regime in $3+1$ dimensions has a discrete spectrum and
its eigenvalues and mass gap are bounded by those of
$\frac{g^{2/3}}{V_{m}^{1/3}}H_{1}$.
\section{Conclusions}
We showed that the most precise condition ensuring the discretness
of the spectrum of the membrane theories and Yang-Mills in the
slow mode regime, is given in terms of an intrinsic moment of
inertia of the membrane. It may be interpreted as if the membrane,
or equivalently the $D0$ branes describing it, have a rotational
energy. It is a quantum mechanical effect. The condition is
obtained from the Molchanov, Maz'ya and Shubin necessary and
sufficient condition on the potential of a Schr\"odinger operator
to have a discrete spectrum. The criteria is expressed not in
terms of the behaviour of the potentials at each point, but by a
mean value, on the configuration space.  The mean value in the
sense of Molchanov considers the integral of the potential on a
finite region of configuration space. It can be naturally
associated to a discretization of configuration space in the
quantum theory. We found that the mean value in the direction of
the valleys where the potential is zero, at large distances in the
configuration space, is the same as a harmonic oscillator with
frequencies given by the tensor of inertia of the membrane.  The
interesting feature is that all previously known bounds for the membrane and
Yang-Mills potential were linear on the configuration variables, while our bound is quadratic on the configuration variables. The bounds we
obtained should also be relevant on the analysis of the spectrum
of wrapped supermembrane with nontrivial central charges
(\cite{br},\cite{bgmr},\cite{bgmmr},\cite{gmr}).
\newline
We also obtained bounds, based on the moment of inertia for the
hamiltonian of Yang-Mills theories in the slow mode regime which
allows to obtain interpretation about the mass gap of the theory.
In particular a bound for $SU(3)$ in $3+1$ dimensions is given by
a hamiltonian whose spectrum and eigenvalues are known, and its
eigenfunctions are expressed in terms of Bessel functions.

In section 3 we obtained a class of potentials, of any degree in
the configuration coordinates, presenting the same valley
properties of the quartic potentials discussed in section 4, 5 and
6. In particular, one of them has similarities with the potential
for the 5-brane in $D=11$. We proved that the associated
Schr\"odinger operator has discrete spectrum, with finite
multiplicity.
\section*{Acknowledgments}
We are very grateful to L. Boulton, F. Cachazo, H. Nicolai, Sh.
Matsuura, V. Strauss for helpful discussions and comments; to C. Burgess, F.
Cachazo, R. Myers, F. Quevedo, for their great support and
encouragment. M.P.G.M. and A.R. would like to thank Perimeter
Institute for kind hospitality while part of this work was done.
M.P.G.M acknowledge support from NSERC of Canada and MEDT of
Ontario, and NSERC Discovery grant.


\begin{thebibliography}{9}
\bibitem{dwln}B. de Wit , M.
Luscher , H. Nicolai {\em The Supermembrane Is Unstable}.
Nucl.Phys.{\bf B320}:135,1989

\bibitem{helling}H Nicolai, R Helling {\em Supermembranes and M(atrix) theory.}
{\tt hep-th/9809103}

\bibitem{simons} B. Simon, Ann. Phys 146 (1983)209

\bibitem{lucher} M. L\"uscher {\emph Some Analytic Results Concerning The Mass Spectrum Of Yang-Mills
Gauge Theories On A Torus}. Nucl.Phys.{\bf B219}:233-261,1983

\bibitem{molchanov}A.M.Molchanov
 {\em On the discretness of the spectrum conditions for selfadjoint differential equations of the second order},
Proc. Moscow Math. Society {\bf 2} (1953) 169-199 (Russian).


\bibitem {Maz-Shu} Mazy'a, V. and Shubin, M. Discreteness of spectrum
and positivity criteria for \schr\ operators. Annals of
Mathematics, \textbf{162} (2005), 919-942.

\bibitem{dwnh} B. de Wit, J. Hoppe, H. Nicolai {\em On The Quantum Mechanics Of
Supermembranes} Nucl.Phys. {\bf B305}:545,1988.

\bibitem{halpern}M. Claudson, M. B. Halpern {\em Supersymmetric
Ground State Wave
  Functions.} Nucl.Phys.{\bf B250}:689,1985

\bibitem{dwmn} B. de Wit, Marquard and H. Nicolai,{\em Area Preserving Diffeomorphisms And Supermembrane Lorentz
Invariance} Commun.Math.Phys. {\bf 128} (1990)39-62.

\bibitem{gabadadze}G. Gabadadze {\em Modeling The Glueball Spectrum By A Closed Bosonic
Membrane} Phys.Rev.{\bf D58}(1998)094015 {\tt hep-ph/9710402}

\bibitem{hoppe}J. Hoppe, MIT Ph.D Thesis, 1982.

\bibitem{l1} L.Boulton, M.P.Garcia del Moral and A. Restuccia. In
preparation.

\bibitem{br} J. Bellorin, A. Restuccia {\tt hep-th/0510259}, {\em D=11 Supermembrane wrapped on
calibrated submanifolds} Nucl.Phys. {\bf B737} (2006) 190-208



 \bibitem{bgmr}L. Boulton, M.P.
Garcia del Moral, A. Restuccia {\em Discreteness of the spectrum
of the compactified D=11 supermembrane with non-trivial winding}
Nucl.Phys. {\bf B671} (2003) 343-358, {\tt hep-th/0211047 }

\bibitem{bgmmr}L. S. Boulton, M. P. Garcia del Moral, I.
Martin, A. Restuccia  {\em On the spectrum of a matrix model for
the D=11 supermembrane compactified on a torus with non-trivial
winding} Class.Quant.Grav. {\bf 19} (2002) 2951 {\tt
hep-th/0109153}

\bibitem{gmr} M.P.Garcia del Moral, A. Restuccia {\em On The
spectrum of a Noncommutative Formulation of the D=11 Supermembrane
with Winding} Phys.Rev. D66 (2002) 045023 {\tt hep-th/0103261}




\end{thebibliography}
\end{document}